\documentclass{ametsocV5}

\usepackage{amsmath,amsfonts,amssymb,bm}
\usepackage{mathptmx}
\usepackage{newtxtext}
\usepackage{newtxmath}
\usepackage{comment}
\usepackage{xcolor}

\title{Using local dynamics to explain analog forecasting of chaotic systems}

\authors{P. Platzer\correspondingauthor{Paul Platzer, paul.platzer@imt-atlantique.fr}}

\affiliation{Laboratoire des Sciences du Climat et de l'Environnement, Saclay, France \& IMT Atlantique, Lab-STICC, UMR CNRS 6285, F-29238, Plouzan\'{e}, France \& France \'{E}nergies Marines, Plouzan\'{e}, France}

\extraauthor{P. Yiou and P. Naveau}
\extraaffil{Laboratoire des Sciences du Climat et de l'Environnement, Saclay, France}

\extraauthor{P. Tandeo and Y. Zhen}
\extraaffil{IMT Atlantique, Lab-STICC, UMR CNRS 6285, F-29238, Plouzan\'{e}, France}

\extraauthor{P. Ailliot}
\extraaffil{Laboratoire de Math\'{e}matiques de Bretagne Atlantique, Brest, France}

\extraauthor{J-F. Filipot}
\extraaffil{France \'{E}nergies Marines, Plouzan\'{e}, France}

\abstract{Analogs are nearest neighbors of the state of a system. By using analogs and their successors in time, one is able to produce empirical forecasts. Several analog forecasting methods have been used in atmospheric applications and tested on well-known dynamical systems. Although efficient in practice, theoretical connections between analog methods and dynamical systems have been overlooked. Analog forecasting can be related to the real dynamical equations of the system of interest. This study investigates the properties of different analog forecasting strategies by taking local approximations of the system's dynamics. We find that analog forecasting performances are highly linked to the local Jacobian matrix of the flow map, and that analog forecasting combined with linear regression allows to capture projections of this Jacobian matrix. The proposed methodology allows to estimate analog forecasting errors, and to compare different analog methods. These results are derived analytically and tested numerically on two simple chaotic dynamical systems.}

\begin{document}


\maketitle

%








\section*{Introduction}

To evaluate the future state of a physical system, one strategy is to use physical knowledge to build differential equations that emulate the dynamics of this system. Then, measurements provide information on the initial state from which these equations must be integrated. Data assimilation gives a framework to account for two main types of error in this forecasting process. First, the aforementioned equations do not describe perfectly the real dynamics of the system, and solving these equations often requires additional approximations, such as numerical discretization. These first error sources combine into what is called model error. Second, observations are usually partial and noisy, such that the initial state from which the differential equations must be integrated is uncertain. Observation error is especially important for chaotic dynamical systems as the latter are highly sensitive to initial conditions.

For complex, highly nonlinear systems such as the atmosphere, forecasts based on physical equations are challenging. Therefore, many empirical methods have been used in atmospheric sciences (see \citealt{van2007empirical}, and references therein). The last decades have seen a proliferation of data from numerical model outputs, observations or the combination of them (see for instance \citealt{saha2010ncep, hersbach2020era5}), strengthening scientific interest for empirical methods. One of such methods is called analog forecasting and is based on a notion originally introduced by \cite{Lorenz1969} to estimate atmospheric predictability. Analog forecasting has been used in meteorological applications and on famous low-dimensional dynamical systems. \cite{Yiou2014} uses analogs in the context of stochastic weather generators. \cite{Tandeo2015}, \cite{Hamilton2016} and \cite{Lguensat2017} combine analog forecasting and data assimilation. More generally, analog forecasting procedures are used in a large range of environmental applications, from tropical intraseasonal oscillations \citep{Alexander2017} to solar irradiance \citep{Ayet2018}.

Analog forecasting proposes to bypass physical equations and to use existing trajectories of the system instead, drawing either from numerical model output, observation data or reanalysis. Analog methods are based on the hypothesis that one is provided with many (or one long) trajectories of the system of interest, which enables to find analog states close to any initial state, and to use the time-successors of these analogs to evaluate the future state of the system. The fluctuating quality and density of available trajectories adds variability to this process. This leads to analog forecasting errors, which play the same role as the previously described model errors.

Preliminary results suggest that analog forecasting errors can be estimated empirically using local approximations of the true dynamics \citep{Platzer2019}. The current paper gives a more in-depth description of the theory that supports different analog forecasting procedures, and allows to evaluate the evolution of analog covariance matrices. The methodology is applied to two famous chaotic Lorenz systems.

The theoretical framework for analog forecasting is outlined in Sec. \ref{sec:AnaFor}, and three analog forecasting operators are recalled. The point of view of dynamical systems is then detailed in Sec. \ref{sec:DynSys}. Finally, Sec. \ref{sec:ConseqAna} examines analog forecasting mean and covariance, and investigates the link between linear regression in analog forecasting and the Jacobian matrix of the real system's flow map. The discussion section takes a broader view, outlines limitations which provide opportunities for new research. The conclusion emphasizes the major results of the paper.

\section{Analog forecasting}\label{sec:AnaFor}

\subsection{Mathematical framework}\label{subsec:mathframe}

Let a dynamical system be defined by the following time-differential equation:

\begin{equation}\label{eq:DSdiff}
    \dfrac{\mathrm{d}\mathbf{x}}{\mathrm{d}t} = \mathbf{f}(\mathbf{x}) \, ,
\end{equation}

\noindent where $\mathbf{x}$ is a vector that fully characterizes the state of the system, and $\mathbf{f}$ is a deterministic, vector-valued map. The space $\mathcal{P}$ in which $\mathbf{x}$ lives is called phase-space. In most applications and throughout this study, $\mathcal{P}$ is a vector space of finite dimension $n$. The system is supposed to be autonomous, such that $\mathbf{f}:\mathcal{P}\rightarrow \mathcal{P}$ does not depend on time.

Given an initial state $\mathbf{x}_0$, a forecast gives an estimation of the state of the system $\mathbf{x}_t$ at a later time $t$. The true future state $\mathbf{x}_t$ is given by the flow map $\boldsymbol{\Phi}:\mathcal{P} \times \mathbb{R} \rightarrow \mathcal{P}$ such that:

\begin{equation}\label{eq:DSflow}
    \boldsymbol{\Phi}^t:\mathbf{x}_0 \rightarrow \boldsymbol{\Phi}^t(\mathbf{x}_0) = \mathbf{x}_t \, .
\end{equation}

For the dynamical system defined through Eq. (\ref{eq:DSdiff}), $\boldsymbol{\Phi}$ represents the time-integration of this equation. For ergodic systems, trajectories come back infinitely close to their initial condition after a sufficiently long time \citep{poincare1890probleme}. Furthermore, if the dynamical system has an attractor set $\mathcal{A}\in \mathcal{P}$, then all trajectories converge to this subset of the phase-space \citep{milnor1985concept}. Analog methods are based on the idea that if one is provided with a long enough trajectory of the system of interest, one will find analog states close to any initial point $\mathbf{x}_0$ in the attractor $\mathcal{A}$. The trajectory from which the analogs are taken is called the "catalog" $\mathcal{C}$, and can either come from numerical model output or reprocessed observational data.

Analog forecasting thus supposes that we know a finite number of initial states that are close enough to $\mathbf{x}_0$ to be called "analogs", and that the flow map of the analogs resembles $\boldsymbol{\Phi}$. Therefore, the time-successors of the analogs should allow to estimate the real future state $\mathbf{x}_t$. In the following, the $k$-th analog and its successor are noted  $\mathbf{a}_0^k$ and $\mathbf{a}_t^k$. Note that analog forecasting is intrinsically random as it depends on the catalog, which is one out of many possible trajectories. The variability in the catalog influences the ability of the analogs and successors to estimate the future state. This motivates the use of probabilistic analog forecasting operators $\boldsymbol{\Theta}$ such that:

\begin{equation}
    \boldsymbol{\Theta}^t : \mathbf{x}_0 \rightarrow \boldsymbol{\Theta}^t ( \mathbf{x}_0 )
\end{equation}

\noindent where $\boldsymbol{\Theta}^t ( \mathbf{x}_0 )$ is a distribution that gives information both about the estimation of the future state $\mathbf{x}_t$ and the variability of this estimation process.

Note that for chaotic dynamical systems, analog forecasting can only work if $t$ is smaller than what is called the "Lyapunov time". This is the characteristic timescale after which trajectories of chaotic systems diverge, such that even if the analog $\mathbf{a}_0^k$ is infinitesimally close to $\mathbf{x}_0$ and if it follows exactly the same dynamics as the real state, the successor $\mathbf{a}_t^k$ will still be far away from $\mathbf{x}_t$. This study is devoted to the properties of analog forecasting below the Lyapunov timescale.

\subsection{Analog forecasting operators}\label{subsec:anaop}

Here are recalled three analog forecasting operators originally introduced in \cite{Lguensat2017}. A finite number $K$ of analogs $(\mathbf{a}_0^k)_{k\in[1,K]}$ and successors $(\mathbf{a}_t^k)_{k\in[1,K]}$ are used, and are assigned weights $(\omega_k)_{k\in[1,K]}$. This allows to give more weight to the pairs of analogs and successors that are best suited for the estimation of $\mathbf{x}_t$. The present article studies the properties of analog forecasting without restriction to any particular choice of weights and distance.\medskip

The distributions of each analog forecast $\boldsymbol{\Theta}^t(\mathbf{x}_0)$ is multinomial, with each pair of analog/successor defining an element of the empirical distribution.

\begin{description}
\item[The locally-constant operator] (LC) uses only the successors to estimate $\mathbf{x}_t$.\\
$\boldsymbol{\Theta}^t_\mathrm{LC}(\mathbf{x}_0) \sim \sum_k \omega_k\boldsymbol{\delta}_{\mathbf{a}_t^k}(\cdot)$ . The mean forecast is thus $\boldsymbol{\mu}_\mathrm{LC} = \sum_k\omega_k\mathbf{a}^k_t$. The covariance of the forecast is $\mathrm{cov}_{\omega_k}(\mathbf{a}_t^k)$, the $\omega$-weighted empirical covariance of the successors.
\end{description}

\begin{description}
\item[The locally-incremental operator] (LI) uses $\mathbf{x}_0$, the analogs and the successors to estimate $\mathbf{x}_t$. \\
$\boldsymbol{\Theta}^t_\mathrm{LI}(\mathbf{x}_0) \sim \sum_k \omega_k\boldsymbol{\delta}_{\mathbf{x}_0+\mathbf{a}_t^k-\mathbf{a}_0^k}(\cdot)$ . The mean forecast is  $\boldsymbol{\mu}_\mathrm{LI} = \mathbf{x}_0 + \sum_k\omega_k(\mathbf{a}^k_t-\mathbf{a}^k_0)$. The covariance of the forecast is $\mathrm{cov}_{\omega_k}(\mathbf{a}^k_t-\mathbf{a}^k_0)$, the $\omega$-weighted empirical covariance of the increments.
\end{description}

\begin{description}
\item[The locally-linear operator] (LL) performs a weighted linear regression between the analogs and the successors. The regression is applied between $\mathbf{a}_0^k-\boldsymbol{\mu}_0$ and the successors $\mathbf{a}_t^k$, where $\boldsymbol{\mu}_0=\sum_k\omega_k\mathbf{a}_0^k$. This gives slope $\mathbf{S}$, intercept $\mathbf{c}$, and residuals $\boldsymbol{\xi}^k=\mathbf{a}_t^k-\mathbf{S}\left(\mathbf{a}_0^k-\boldsymbol{\mu}_0\right)-\mathbf{c}$. \\
 $\boldsymbol{\Theta}^t_\mathrm{LL}(\mathbf{x}_0) \sim \sum_k \omega_k\boldsymbol{\delta}_{\boldsymbol{\mu}_\mathrm{LL}+\boldsymbol{\xi}^k}(\cdot)$ . The mean forecast is $\boldsymbol{\mu}_\mathrm{LL} = \mathbf{S}\left(\mathbf{x}_0-\boldsymbol{\mu}_0\right) + \mathbf{c}$. The covariance of the forecast is $\mathrm{cov}_{\omega_k}(\boldsymbol{\xi}_k)$, the $\omega$-weighted empirical covariance of the residuals.
\end{description}

\begin{figure}[t]
    \noindent\includegraphics[scale=.67]{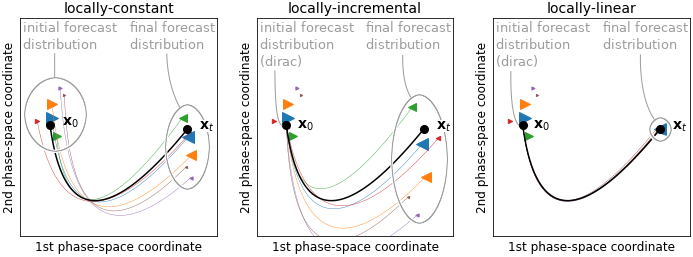}
    \caption{Analog forecasting operators presented in Sec. \ref{subsec:anaop}. The flow map $\boldsymbol{\Phi}^t(\mathbf{x}_0)$ has a simple polynomial form. Analogs are drawn from a normal distribution centered on $\mathbf{x}_0$ and follow the same model as the real state $\mathbf{x}$. The same analogs and flow maps are used for the three operators and are represented on each panel. Weights $\omega_k$ are computed using Gaussian kernels. The real initial and future states $\mathbf{x}_0$ and $\mathbf{x}_t$ are displayed in full circles. On the left panel, analogs are in colored, right-pointing triangles, and successors in left-pointing triangles with the same colors. The size of the $k$-th triangle is proportional to the weight $\omega_k$. In the middle and right panels, the elements of the forecast distribution at time $t$ are also in colored, left-pointing triangles. }
    \label{fig:afop}
\end{figure}

The locally-constant (LC), locally-incremental (LI) and locally-linear (LL) analog forecasting operators are illustrated in Fig. \ref{fig:afop}. The variance of the LC is similar around $t=0$ and for the final value of $t$. On the other hand, the variance of the LI goes to 0 as $t\rightarrow 0$, but for large times the LI estimator has a larger variance compared to the LC. The next sections provide some information that help interpreting this phenomenon. The LL is able to catch the dynamics, and therefore shows a small variance and a good precision at all times. This is due to the fact that, in this example, non-linear terms are small and the flow map of the analogs matches exactly the real system's flow map.

It is worth mentioning another kind of analog forecasting operator called "constructed analogs" (CA). It is a particular case of the locally-constant operator where the weights $\omega_k^\mathrm{CA}$ can have negative values and are such that the mean of the analogs $\boldsymbol{\mu}_0$ is as close as possible to the initial state: $\left\lbrace \omega_k^\mathrm{CA} \right\rbrace_k = \mathrm{argmin}_{\lbrace\omega_k\rbrace_k} |\sum_k\omega_k\mathbf{a}_0^k-\mathbf{x}_0|$. It was used by \cite{VanDenDool1994} to create better analogs in the case of small catalogs. Later, \cite{Tipett2013} showed that CA are equivalent to the locally-linear operator with constant weights. In the following and unless otherwise specified, it is assumed that the weights $\omega_k$ are positive and decreasing functions of the distance between $\mathbf{a}_0^k$ and $\mathbf{x}_0$.

\section{Successor-to-future state distance}\label{sec:DynSys}

\subsection{Notations and hypotheses}

This work assumes that the evolution dynamics of the analogs are similar to the evolution dynamics of the system of interest, and that the system is deterministic. This can be stated in a differential equation form:

\begin{subequations}\label{eq:Dynf}
\begin{equation}
\begin{cases} \dfrac{\mathrm{d}\mathbf{x}}{\mathrm{d}t}=
               \mathbf{f}(\mathbf{x})\\
               \mathbf{x}_{t=0}=\mathbf{x}_0\\
            \end{cases} \, ,
\end{equation}
\begin{equation}
\forall k,\, \begin{cases} \dfrac{\mathrm{d}\mathbf{a}^k}{\mathrm{d}t}=
               \mathbf{f}_a(\mathbf{a}^k)\\
               \mathbf{a}_{t=0}^k=\mathbf{a}_0^k\\
            \end{cases} \; , \quad \mathrm{with} \quad \mathbf{f}_a=\mathbf{f}+\delta \tilde{\mathbf{f}} \, ,
\end{equation}
\end{subequations}\medskip

\noindent or in an integrated form using flow maps:
 
\begin{subequations}\label{eq:DynPhi}
\begin{equation}
\mathbf{x}_t=\boldsymbol{\Phi}^t(\mathbf{x}_0) \, ,
\end{equation}
\begin{equation}
\forall k,\, \mathbf{a}^k_t=\boldsymbol{\Phi}^t_a(\mathbf{a}^k_0) \; , \quad \mathrm{with} \quad \boldsymbol{\Phi}^t_a=\boldsymbol{\Phi}^t+\delta \tilde{\boldsymbol{\Phi}}^t \, ,
\end{equation}
\end{subequations}

\noindent where $\boldsymbol{\Phi}_a$ is the flow map of the analogs, and $\tilde{\boldsymbol{\Phi}}$ is the difference between the analog and real flow maps normalized through the scalar value $\delta$ such that $\boldsymbol{\Phi}$, $\boldsymbol{\Phi}_a$ and $\tilde{\boldsymbol{\Phi}}$ are of the same order of magnitude. The maps $\mathbf{f}$, $\mathbf{f}_a$ and $\tilde{\mathbf{f}}$ are defined accordingly. 

\medskip

In these formulations, the fundamental hypotheses of analog forecasting are the continuity of $\boldsymbol{\Phi}^t$ (or $\mathbf{f}$) with respect to the phase-space variable $\mathbf{x}$, the density of the catalog $\mathcal{C}$ (for all $k$, $\mathbf{a}_0^k$ is close to $\mathbf{x}_0$ for a given metric) and the adequacy of the analogs' dynamics ($\delta$ is small, $\boldsymbol{\Phi}_a\approx \boldsymbol{\Phi}$).

The next section will investigate the ability of analogs and successors to approximate the real system state, provided that $t$ is below the Lyapunov time and that the aforementioned hypotheses are verified.

\subsection{When analogs work : Taylor expansions of the dynamics} \label{subsec:TayloAna}

\subsubsection{Distance between successor and real state}

\begin{figure}
    \noindent\includegraphics[scale=.54]{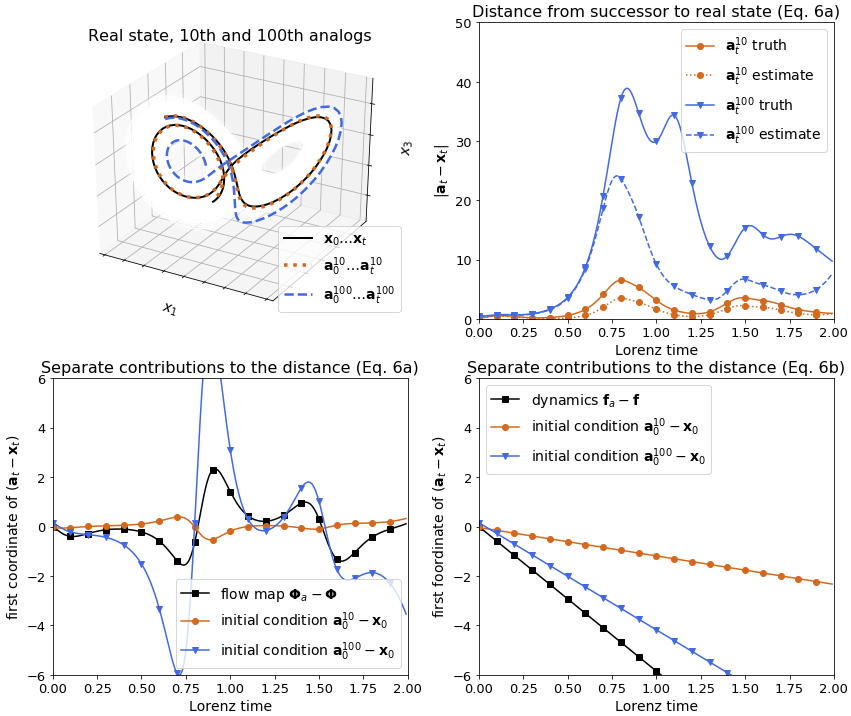}
    \caption{Illustrating Eq. (\ref{eq:Taylor_Ana}a,b) on the three-variable L63 system. Upper-left: A real trajectory from $\mathbf{x}_0$ to $\mathbf{x}_t$ and two analog trajectories, namely the 10-th best analog $\mathbf{a}^{10}_0$ to $\mathbf{a}^{10}_t$ and the 100-th best analog $\mathbf{a}^{100}_0$ to $\mathbf{a}^{100}_t$. The catalog is shown in white. Upper-right: comparing the exact value of the norm of $\mathbf{a}_t-\mathbf{x}_t$ (full lines) and the sum of the two terms on the right-hand side of Eq. (\ref{eq:Taylor_Ana_Phi}) (dashed lines). Lower-left: Contributions of the first term (black squares) and the second term (brown circles and blue triangles) of the right-hand side of equation (\ref{eq:Taylor_Ana_Phi}) projected on the first coordinate of the L63 system. Lower-right: Contributions of the first term (black squares) and the second term (brown circles and blue triangles) of the right-hand side of equation (\ref{eq:Taylor_Ana_f_Euler}) projected on the first coordinate of the L63 system.}
    \label{fig:validity_lin}
\end{figure}

Assuming different levels of smoothness of the flow maps and using Taylor expansions, one can estimate the difference between the real future state $\mathbf{x}_t$ and any given successor $\mathbf{a}_t^k$ at leading order:

\begin{subequations}\label{eq:Taylor_Ana}
\begin{equation}\label{eq:Taylor_Ana_Phi}
    \forall k, \quad \mathbf{a}_t^k-\mathbf{x}_t= \delta \tilde{\boldsymbol{\Phi}}^t(\mathbf{x}_0)+ \left[\boldsymbol{\nabla}\boldsymbol{\Phi}^t|_{\mathbf{x}_0}\right]\cdot(\mathbf{a}^k_0-\mathbf{x}_0)+\mathcal{O}\left( |\mathbf{a}^k_0-\mathbf{x}_0|^2 \, , \; \delta |\mathbf{a}^k_0-\mathbf{x}_0|  \right) \, ,
\end{equation}

\noindent where $\boldsymbol{\nabla}\boldsymbol{\Phi}^t|_{\mathbf{x}_0}$ is the Jacobian matrix (the matrix of partial derivatives in phase-space) of $\boldsymbol{\Phi}^t$ at $\mathbf{x}_0$, '$\cdot$' is the matrix multiplication, and $\mathcal{O}\left( |\mathbf{a}^k_0-\mathbf{x}_0|^2 \, , \; \delta |\mathbf{a}^k_0-\mathbf{x}_0|  \right)$ represents higher-order terms. Neglecting these higher-order terms and lightening notations, this equation can be rewritten:

\begin{equation*}
    \mathbf{a}_t^k-\mathbf{x}_t \approx \delta \tilde{\boldsymbol{\Phi}}^t+ \boldsymbol{\nabla}\boldsymbol{\Phi}^t \cdot(\mathbf{a}^k_0-\mathbf{x}_0) \, ,
\end{equation*}

\noindent where the evaluation of $\delta \tilde{\boldsymbol{\Phi}}^t$ and $\boldsymbol{\nabla}\boldsymbol{\Phi}^t$ at $\mathbf{x}_0$ is implicit. The leading-order difference terms explicitly described in the right hand-side of Eq. (\ref{eq:Taylor_Ana_Phi}) come from two sources. The first source is the difference between the analog and true flow maps at point $\mathbf{x}_0$, which is independent of $\mathbf{a}_0^k$. The second source of difference is the mismatch in the initial condition, left-multiplied by the Jacobian matrix of the true flow map at point $\mathbf{x}_0$.

Eq. (\ref{eq:Taylor_Ana_Phi}) states that at first order, these two error terms are additive. This is not true at higher orders. Higher-order terms include the bilinear product of $\mathbf{a}^k_0-\mathbf{x}_0$ with a matrix of second derivatives of $\boldsymbol{\Phi}^t$ called the Hessian, and the product of the Jacobian of $\tilde{\boldsymbol{\Phi}}^t$ at $\mathbf{x}_0$ and $\mathbf{a}^k_0-\mathbf{x}_0$.

Fig. \ref{fig:validity_lin} shows applications of Eq. (\ref{eq:Taylor_Ana_Phi}) to the three-variable system of \cite{Lorenz1963}, hereafter noted L63. A real trajectory is compared with two analog trajectories. The L63 system is solved numerically using a fourth-order Runge-Kutta finite-difference scheme, with numerical integration time step $\Delta t=0.01$ non-dimensional time. For notation details, see Eq. (\ref{eq:L63}) in appendix \ref{app:lorenz}. The real trajectory has parameters $\sigma=10$, $\rho=28$, $\beta=8/3$, while the $\sigma$ parameter for the analog dynamics is slightly perturbed with $\sigma_a=9=0.9\sigma$. The matrices $\delta \tilde{\boldsymbol{\Phi}}^t$ and $\boldsymbol{\nabla}\boldsymbol{\Phi}^t$ are estimated numerically using formulae given below and time step $\Delta t=0.01$. The 10-th analog stays close enough to the real trajectory all the time (upper-left panel of Fig. \ref{fig:validity_lin}), therefore Eq. (\ref{eq:Taylor_Ana_Phi}) gives a satisfactory approximation of $|\mathbf{a}^{10}_t-\mathbf{x}_t|$ (upper-right panel). The 100-th analog 
starts to be too far from the real trajectory around $t\approx0.7$ (upper-left and right panels), and  Eq. (\ref{eq:Taylor_Ana_Phi}) provides a poor approximation of $|\mathbf{a}^{100}_t-\mathbf{x}_t|$ (upper-right panel).

The different right-hand side-terms of Eq. (\ref{eq:Taylor_Ana_Phi}) are projected on the first axis of phase-space and displayed in the lower-left panel of Fig. \ref{fig:validity_lin}. The "flow map" term $\delta\tilde{\boldsymbol{\Phi}}^t$ is the same for both analogs, but the "initial condition" term $\boldsymbol{\nabla}\boldsymbol{\Phi}^t \cdot (\mathbf{a}_0^k-\mathbf{x}_0)$ is much larger for the 100-th analog, and one can see that those terms are proportional, here negatively correlated. \medskip

Further assuming that $t$ is small, one can express Eq. (\ref{eq:Taylor_Ana_Phi}) in the alternative formulation:

\begin{equation}\label{eq:Taylor_Ana_f_Euler}
    \forall k, \quad \mathbf{a}_t^k-\mathbf{x}_t= t \delta \tilde{\mathbf{f}}(\mathbf{x}_0)+ \left[ \mathbf{I} + t \boldsymbol{\nabla}\mathbf{f}|_{\mathbf{x}_0} \right]\cdot (\mathbf{a}^k_0-\mathbf{x}_0)+\mathcal{O}\left(t^2 \, , \; |\mathbf{a}^k_0-\mathbf{x}_0|^2 \, , \; \delta |\mathbf{a}^k_0-\mathbf{x}_0|  \right) \, ,
\end{equation}
\end{subequations}

\noindent where $\mathbf{I}$ is the identity matrix. Using lighter notations, this becomes:

\begin{equation*}
    \mathbf{a}_t^k-\mathbf{x}_t \approx t \delta \tilde{\mathbf{f}}+ \left[ \mathbf{I} + t \boldsymbol{\nabla}\mathbf{f} \right]\cdot (\mathbf{a}^k_0-\mathbf{x}_0) \, ,
\end{equation*}

\noindent where the evaluation of $\delta \tilde{\mathbf{f}}$ and $\boldsymbol{\nabla}\mathbf{f}$ at $\mathbf{x}_0$ is implicit. This last formulation is analogous to a Euler scheme used in finite-difference numerical methods for solving differential equations, it is therefore valid only for small times. In the lower-right panel of Fig. \ref{fig:validity_lin}, one can see that the right-hand side terms of Eq. (\ref{eq:Taylor_Ana_f_Euler}) only approximate the terms of Eq. (\ref{eq:Taylor_Ana_Phi}) for $t\lesssim 0.1$.\medskip

\subsubsection{Link between the two formulations, $\mathbf{f}$ and $\boldsymbol{\Phi}^t$}
\label{subsec:link}

Eq. (\ref{eq:Taylor_Ana_f_Euler}) is a first-order expansion in time of Eq. (\ref{eq:Taylor_Ana_Phi}) . The fundamental resolvent matrix $\mathbf{M}(t,t')$ gives a more complete relationship between the two representations. $\mathbf{M}(t,t')$ is solution to the time-varying linear system $\frac{\mathrm{d}\mathbf{M}(t,t')}{\mathrm{d}t} = \boldsymbol{\nabla}\mathbf{f}|_{\mathbf{x}_t} \cdot \mathbf{M}(t,t')$ with $\mathbf{M}(t',t')=\mathbf{I}$. The fundamental resolvent matrix can be approximated numerically as $\mathbf{M}(t,t')\approx \exp(\Delta t\mathbf{\nabla}\mathbf{f}_t)\cdot \exp(\Delta t\mathbf{\nabla}\mathbf{f}_{t-\Delta t}) \dots \exp(\Delta t\mathbf{\nabla}\mathbf{f}_{t'})$ with numerical time-step $\Delta t$ and using the short notation $\mathbf{\nabla}\mathbf{f}_t:=\mathbf{\nabla}\mathbf{f}|_{\mathbf{x}_t}$.

We have:

\begin{subequations}\label{eq:Phitof}
\begin{equation}
\delta \tilde{\boldsymbol{\Phi}}^t(\mathbf{x}_0) \approx \delta\int_0^t \mathbf{M}(t,t') \cdot  \tilde{\mathbf{f}}(\mathbf{x}_{t'}) \,\mathrm{d}u \, ,
\end{equation}
\begin{equation}
\boldsymbol{\nabla}\boldsymbol{\Phi}^t|_{\mathbf{x}_0} = \mathbf{M}(t,0)\, ,
\end{equation}
\end{subequations}

\noindent where the "$\approx$" sign is here to say that Eq. (\ref{eq:Phitof}a) is valid only at first order in $\delta$. This first order is enough to compute the right-hand side terms of Eq. (\ref{eq:Taylor_Ana_Phi}), which is also valid at first order in $\delta$. 

From Eq. (\ref{eq:Phitof}b) one can use Taylor developments relating $\boldsymbol{\nabla}\mathbf{f}$ and $\boldsymbol{\nabla}\boldsymbol{\Phi}^t$, such as:

\begin{equation}\label{eq:Phitof_tsquared}
\boldsymbol{\nabla}\boldsymbol{\Phi}^t = \mathbf{I} + t\, \boldsymbol{\nabla}\mathbf{f}_0 + t^2 \, \left( \left(\boldsymbol{\nabla}\mathbf{f}_0\right)^2 + \frac{\mathrm{d}}{\mathrm{d}t}\boldsymbol{\nabla}\mathbf{f}_0 \right) + \mathcal{O}(t^3) \, ,
\end{equation}
 
\noindent where $\boldsymbol{\nabla}\boldsymbol{\Phi}^t$ is implicitly evaluated at $\mathbf{x}_0$. The short notation $\boldsymbol{\nabla}\mathbf{f}_0$ is used for $\boldsymbol{\nabla}\mathbf{f}_{\mathbf{x}_0}$, and  $\frac{\mathrm{d}}{\mathrm{d}t}\boldsymbol{\nabla}\mathbf{f}_0$ is the time derivative along the trajectory $\mathbf{x}_t$ of the Jacobian of $\mathbf{f}$, at $t=0$. $\frac{\mathrm{d}}{\mathrm{d}t}\boldsymbol{\nabla}\mathbf{f}|_0:=\lim_{t\rightarrow 0} \left( \boldsymbol{\nabla}\mathbf{f}_t-\boldsymbol{\nabla}\mathbf{f}_0 \right)/t$. At first order in $t$, one recovers the result expressed in Eq. (\ref{eq:Taylor_Ana_f_Euler}).

\section{Consequences for analog forecasting operators} \label{sec:ConseqAna}

\subsection{Mean error of analog forecasting operators} \label{subsec:MeanAna}

By multiplying equations (\ref{eq:Taylor_Ana}a,b) by $\omega_k$ and summing over $k$, one can compare the distances from $\mathbf{x}_t$ to the averages $\boldsymbol{\mu}_\mathrm{LC}$, $\boldsymbol{\mu}_\mathrm{LI}$ and $\boldsymbol{\mu}_\mathrm{LL}$ of the different analog forecasting operators of Sec. \ref{sec:AnaFor}\ref{subsec:anaop}. Those averages depend on $t$, although only implicitly in the notation. Letting $\boldsymbol{\mu}_0=\sum_k\omega_k\mathbf{a}_0^k$ the weighted mean of the analogs, we have the following expressions.

\begin{subequations}\label{eq:LCmeanERR}
\begin{description}
\item[Locally-constant mean error]:
\begin{equation}
    \boldsymbol{\mu}_\mathrm{LC}-\mathbf{x}_t=\delta \tilde{\boldsymbol{\Phi}}^t(\mathbf{x}_0)+ \left[\boldsymbol{\nabla}\boldsymbol{\Phi}^t|_{\mathbf{x}_0}\right]\cdot(\boldsymbol{\mu}_0-\mathbf{x}_0)+\mathcal{O}\left( \sum_k\omega_k|\mathbf{a}^k_0-\mathbf{x}_0|^2 \, , \; \delta\sum_k\omega_k |\mathbf{a}^k_0-\mathbf{x}_0|  \right) \, , 
\end{equation}
\begin{equation}
    \boldsymbol{\mu}_\mathrm{LC}-\mathbf{x}_t=t\delta \tilde{\mathbf{f}}(\mathbf{x}_0)+ \left[\mathbf{I} + t \boldsymbol{\nabla}\mathbf{f}|_{\mathbf{x}_0} \right]\cdot (\boldsymbol{\mu}_0-\mathbf{x}_0)+\mathcal{O}\left(t^2 \, , \; \sum_k\omega_k|\mathbf{a}^k_0-\mathbf{x}_0|^2 \, , \; \delta \sum_k\omega_k|\mathbf{a}^k_0-\mathbf{x}_0|  \right) \, .
\end{equation}
\end{description}
\end{subequations}

\begin{subequations}\label{eq:LImeanERR}
\begin{description}
\item[Locally-incremental mean error]:
\begin{equation}
    \boldsymbol{\mu}_\mathrm{LI}-\mathbf{x}_t=\delta \tilde{\boldsymbol{\Phi}}^t(\mathbf{x}_0)+ \left[\boldsymbol{\nabla}\boldsymbol{\Phi}^t|_{\mathbf{x}_0}-\mathbf{I}\right]\cdot (\boldsymbol{\mu}_0-\mathbf{x}_0)+\mathcal{O}\left( \sum_k\omega_k|\mathbf{a}^k_0-\mathbf{x}_0|^2 \, , \; \delta \sum_k\omega_k|\mathbf{a}^k_0-\mathbf{x}_0|  \right) \, ,
\end{equation}
\begin{equation}
    \boldsymbol{\mu}_\mathrm{LI}-\mathbf{x}_t = t\delta \tilde{\mathbf{f}}(\mathbf{x}_0)+ \left[t\boldsymbol{\nabla}\mathbf{f}|_{\mathbf{x}_0}\right]\cdot (\boldsymbol{\mu}_0-\mathbf{x}_0)+\mathcal{O}\left( t^2, \sum_k\omega_k|\mathbf{a}^k_0-\mathbf{x}_0|^2 \, , \; \delta \sum_k\omega_k|\mathbf{a}^k_0-\mathbf{x}_0|  \right) \, .
\end{equation}
\end{description}
\end{subequations}

Using lighter notations with implicit evaluation at $\mathbf{x}_0$, this gives:

\begin{align*}
    \mathrm{Locally-constant: }\, \boldsymbol{\mu}_\mathrm{LC}-\mathbf{x}_t \approx &\,  \delta \tilde{\boldsymbol{\Phi}}^t+ \boldsymbol{\nabla}\boldsymbol{\Phi}^t\cdot(\boldsymbol{\mu}_0-\mathbf{x}_0) \\
    \approx & \, t\delta \tilde{\mathbf{f}}+ \left[\mathbf{I} + t  \boldsymbol{\nabla}\mathbf{f} \right]\cdot (\boldsymbol{\mu}_0-\mathbf{x}_0) \, , \\
    \mathrm{Locally-incremental: }\, \boldsymbol{\mu}_\mathrm{LI}-\mathbf{x}_t \approx &\,  \delta \tilde{\boldsymbol{\Phi}}^t+ \left[\boldsymbol{\nabla}\boldsymbol{\Phi}^t-\mathbf{I}\right]\cdot (\boldsymbol{\mu}_0-\mathbf{x}_0) \\
    \approx &\, t\delta \tilde{\mathbf{f}} + t\boldsymbol{\nabla}\mathbf{f}\cdot (\boldsymbol{\mu}_0-\mathbf{x}_0) \, .
\end{align*}

The errors of the locally-constant and locally-incremental operators are both affected by the difference between the analog and real flow maps. This source of error cannot be circumvented unless provided with some information about $\delta\tilde{\boldsymbol{\Phi}}$. The other first-order error term is linear in $(\boldsymbol{\mu}_0-\mathbf{x}_0)$, but when $t\rightarrow 0$, this term is of order $t$ in the locally-incremental case. Thus, for small lead-times, as both $t\rightarrow 0$ and $\boldsymbol{\mu}_0\rightarrow\mathbf{x}_0$ (dense catalog), the mean of the locally-incremental provides a better estimate of $\mathbf{x}_t$. This is why this operator is qualified by \cite{Lguensat2017} as more "physically-sound" than the locally-constant: the locally-incremental takes advantage of the fact that $\lim_{t\rightarrow 0}\boldsymbol{\Phi}^t=\mathbf{I}$, just as any finite-difference numerical scheme does. Formulas similar to Eq. (\ref{eq:LCmeanERR}-\ref{eq:LImeanERR}) were used by \cite{Platzer2019} to predict analog forecasting errors with LC and LI operators, on the famous three-variable L63 system, with $\delta=0$. 

Another interesting property of the locally-incremental is that it can give estimates of $\mathbf{x}_t$ out of the convex hull of the catalog. This is related to what is called "novelty creation" in the machine-learning terminology. Such a property is interesting, but it also enables some inconsistent forecasts. Indeed, if $t$ is not small enough, the locally-incremental operator can produce forecasts that have a large error due to the $-\mathbf{I}$ term in Eq. (\ref{eq:LImeanERR}a). In Fig. \ref{fig:afop}, one can see that the LI has a larger variance than the LC for large times.

Eq. (\ref{eq:LCmeanERR}) is also valid for constructed analogs (CA) introduced in section \ref{subsec:anaop}, where the weights $\lbrace\omega_k^\mathrm{CA}\rbrace$ are chosen so that $|\sum_k\omega_k^\mathrm{CA}\mathbf{a}_0^k-\mathbf{x}_0|$ is as small as possible. This means that the $(\boldsymbol{\mu}_0-\mathbf{x}_0)$-linear term of equation (\ref{eq:LCmeanERR}) is also small. As mentioned earlier, \cite{Tipett2013} showed that this strategy is equivalent to making a linear regression. This explains why the $(\boldsymbol{\mu}_0-\mathbf{x}_0)$-linear term is absent from Eqs. (\ref{eq:LLmeanERR}a,b).

\begin{subequations}\label{eq:LLmeanERR}
\begin{description}
\item[Locally-linear mean error]:
\begin{equation}
\boldsymbol{\mu}_\mathrm{LL}-\mathbf{x}_t = \delta \tilde{\boldsymbol{\Phi}}^t(\mathbf{x}_0)+ \mathcal{O}\left( \sum_k\omega_k|\mathbf{a}^k_0-\mathbf{x}_0|^2 \, , \; \delta \sum_k\omega_k|\mathbf{a}^k_0-\mathbf{x}_0|  \right) \, ,
\end{equation}
\begin{equation}
\boldsymbol{\mu}_\mathrm{LL}-\mathbf{x}_t = t\delta \tilde{\mathbf{f}}(\mathbf{x}_0)+ \mathcal{O}\left( t^2 \, , \; \sum_k\omega_k|\mathbf{a}^k_0-\mathbf{x}_0|^2 \, , \; \delta \sum_k\omega_k|\mathbf{a}^k_0-\mathbf{x}_0|  \right) \, .
\end{equation}
\end{description}
\end{subequations}

Another way to understand why the $(\boldsymbol{\mu}_0-\mathbf{x}_0)$-linear term should disappear when using the LL is to see that the LL is estimating the local Jacobian of the flow map. Indeed, the linear regression between the analogs and the successors gives an estimation of $\boldsymbol{\nabla}\boldsymbol{\Phi}^t|_{\mathbf{x}_0}$, with an estimation error that is at least of order $\mathcal{O}(|\boldsymbol{\mu}_0-\mathbf{x}_0|,\delta)$. Sec. \ref{sec:ConseqAna}.\ref{subsec:commentS} gives a detailed argumentation to support this claim and investigates limitations. The estimation error between the linear regression matrix and the Jacobian thus adds higher-order error terms to the right-hand side of Eqs. (\ref{eq:LLmeanERR}a,b), but these are already included in the $\mathcal{O}(\sum_k\omega_k|\mathbf{a}^k_0-\mathbf{x}_0|^2 \, , \; \delta \sum_k\omega_k|\mathbf{a}^k_0-\mathbf{x}_0|)$.\medskip

We now make the explicit link between the three operators. Recall the notations of Sec. \ref{sec:AnaFor}.\ref{subsec:anaop}: the locally-linear operator finds slope $\mathbf{S}$ and intercept $\mathbf{c}$ such that for all $ k$, $ \mathbf{a}_t^k=\mathbf{S}(\mathbf{a}_0^k-\boldsymbol{\mu}_0)+\mathbf{c}+\boldsymbol{\xi}^k$ using weighted least-square estimates. This gives $\mathbf{c}=\sum_k\omega_k\mathbf{a}_t^k=\boldsymbol{\mu}_\mathrm{LC}$, thus we have $\boldsymbol{\mu}_\mathrm{LL}=\boldsymbol{\mu}_\mathrm{LC}+\mathbf{S}(\mathbf{x}_0-\boldsymbol{\mu}_0)$ and the following relations hold:

\begin{subequations}
\begin{equation}
\boldsymbol{\mu}_\mathrm{LC}=\boldsymbol{\mu}_\mathrm{LL}|_{\mathbf{S}=\mathbf{0}}\, ,
\end{equation}
\begin{equation}
\boldsymbol{\mu}_\mathrm{LI}=\boldsymbol{\mu}_\mathrm{LL}|_{\mathbf{S}=\mathbf{I}}\, ,
\end{equation}
\end{subequations}

\noindent such that the locally-constant and locally-incremental operator are particular cases of the locally-linear operator. We also have $\lim_{t\rightarrow 0}\mathbf{S}=\mathbf{I}$, because for all $k$, $\lim_{t\rightarrow 0} \mathbf{a}^k_t=\mathbf{a}^k_0$. Thus, mean forecasts of the locally-linear and locally-incremental operators are equivalent as $t$ approaches 0: $\boldsymbol{\mu}_\mathrm{LL} \sim_{t\rightarrow 0} \boldsymbol{\mu}_\mathrm{LI}$.\medskip

This analysis shows that, in terms of mean forecast error, the locally-linear operator is more precise than the locally-incremental, and the latter is more precise than the locally-constant. These findings are in agreement with the numerical experiments of \cite{Lguensat2017}.

We now investigate  the link between the local Jacobian of the flow $\boldsymbol{\nabla}\boldsymbol{\Phi}^t|_{\mathbf{x}_0}$, and the linear regression matrix from the locally-linear operator $\mathbf{S}$.

\subsection{Ability of analogs to estimate local Jacobians}\label{subsec:commentS}

If analogs can estimate the Jacobian of the real system, it means that analog forecasting provides a local approximation of the real dynamics, proving the relevance of analogs for short-range forecasts. Furthermore, having an estimation of the local Jacobian can be useful in some applications such as the Extend Kalman Filter, where the Jacobian allows to estimate the evolution of the forecast covariance.

\subsubsection{Derivation of the first order error in Jacobian estimation}

It is possible to find an exact expression of the first-order error term in the estimation of the local Jacobian. Let us start with the case of perfect agreement between the real and analog flow maps: $\boldsymbol{\Phi}_a=\boldsymbol{\Phi}$, or $\delta=0$. Then, assume that in the neighborhood of $\mathbf{x}_0$ where the analogs lie, the flow $\boldsymbol{\Phi}^t(\cdot)$ can be approximated by a quadratic function in phase-space. We then have :

\begin{equation}\label{eq:Fquad}
\forall k \, , \quad \mathbf{a}^k_t=\boldsymbol{\nabla}\boldsymbol{\Phi}^t (\mathbf{a}^k_0-\boldsymbol{\mu}_0)+ \frac{1}{2}  (\mathbf{a}^k_0-\boldsymbol{\mu}_0) \boldsymbol{\nabla}^2\boldsymbol{\Phi}^t (\mathbf{a}^k_0-\boldsymbol{\mu}_0)^\mathrm{T}+\mathrm{Cst}\, ,
\end{equation}

\noindent where "Cst" is a constant (independent of $k$), and the Jacobian and Hessian of $\boldsymbol{\Phi}^t$ are implicitly evaluated at $\mathbf{x}_0$ (see appendix \ref{app:hessian} for notation of product of vectors and Hessian). In the next equations, the $t$-superscript is dropped to lighten notations. Let $\mathbf{X}$, the matrix of the analogs minus their mean, so that the $k$-th row of $\mathbf{X}$ is $\mathbf{a}^k_0-\boldsymbol{\mu}_0$. Similarly, let $\mathbf{Y}$ be the matrix of the successors, with the $k$-th row of $\mathbf{Y}$ being $\mathbf{a}^k_t$. Eq. (\ref{eq:Fquad}) thus translates into $\mathbf{Y}=\mathbf{X}\boldsymbol{\nabla}\boldsymbol{\Phi}^\mathrm{T}+\frac{1}{2}\mathbf{X}\boldsymbol{\nabla}^2\boldsymbol{\Phi}\mathbf{X}^\mathrm{T}$, omitting the constant.

Now let $\boldsymbol{\Omega}=\mathrm{diag}(\omega_1, \, \dots,\, \omega_K )$, the $(K\times K)$ diagonal matrix of the weights given to each analog in the regression. Then $\mathbf{S}$ is the weighted least-squares solution of the linear regression $\mathbf{S}=(\mathbf{X}^\mathrm{T}\boldsymbol{\Omega}\mathbf{X})^{-1}\mathbf{X}^\mathrm{T}\boldsymbol{\Omega}\mathbf{Y}$. With a bit of rewriting, this finally gives:

\begin{equation}\label{eq:Slin_mat}
\mathbf{S}-\boldsymbol{\nabla}\boldsymbol{\Phi}=(\mathbf{X}^\mathrm{T}\boldsymbol{\Omega}\mathbf{X})^{-1}\mathbf{X}^\mathrm{T}\boldsymbol{\Omega} \left[  \mathbf{X}\; \boldsymbol{\nabla}^2\boldsymbol{\Phi}\; \left(\frac{1}{2} \mathbf{X} + (\boldsymbol{\mu}_0-\mathbf{x}_0)^\mathrm{T} \otimes \mathbf{J}_{K,1} \right)^\mathrm{T} \right]\, ,
\end{equation}
 
\noindent where $\otimes$ is the Kronecker matrix product and $\mathbf{J}_{K,1}$ is the column vector with $K$ elements all equal to 1.
 
Eq. (\ref{eq:Slin_mat}) tells us that $\mathbf{S}$ is close to the Jacobian at $\mathbf{x}_0$ up to a factor that is linear in the distance between the mean of the analogs $\boldsymbol{\mu}_0$ and the analogs $\mathbf{a}^k_0$, and another factor linear in the distance between $\boldsymbol{\mu}_0$ and $\mathbf{x}_0$. These linear error term depend on the second-order phase-space derivatives of $\boldsymbol{\Phi}$ at the point $\mathbf{x}_0$ (the Hessian of $\boldsymbol{\Phi}$).

Conducting the same derivation but relaxing the hypothesis of $\delta=0$, one would find the same result with an added linear error term involving the Jacobian of $\tilde{\boldsymbol{\Phi}}$. This analysis allows us to say that $\mathbf{S}=\boldsymbol{\nabla}\boldsymbol{\Phi}^t+\mathcal{O}\left( |\boldsymbol{\mu}_0-\mathbf{x}_0|, \, \delta \right)$, if the distance between the analogs and their mean is of same order as the distance between their mean and $\mathbf{x}_0$.

However, the claim that the linear regression matrix $\mathbf{S}$ is able to approximate the Jacobian $\boldsymbol{\nabla}\boldsymbol{\Phi}^t$ must be tempered by several facts. To illustrate these, the regular locally-linear analog forecasting operator will now be compared with two other strategies aimed at solving dimensionality issues.

\subsubsection{Strategies for linear regression in high dimension}

Dimensionality can make analog forecasting difficult, especially when using the locally linear analog forecasting operator. Here are recalled two strategies that can be used to circumvent this issue.

The first approach uses empirical orthogonal functions (EOFs, also called principal component analysis) at every forecast step. Dimension is reduced by keeping only the first $n^\mathrm{eof}$ EOFs of the set of analogs $(\mathbf{a}_0^k)_{k\in[1,K]}$, or keeping only the $n^\mathrm{eof}$ first principal components of the matrix $\mathbf{X}^\mathrm{T}\boldsymbol{\Omega}\mathbf{X}$. \medskip

\textbf{Reducing dimension using EOFs}

\begin{itemize}
\small
    \item Find analogs $(\mathbf{a}^k_0)_{k\in[1,K]}$ of the initial state $\mathbf{x}_0$
    \item Compute the $n$ EOFs of the weighted set of analogs $(\mathbf{a}^k_0)_{k\in[1,K]}$
    \item Keep the $n^\mathrm{eof}$ first EOFs up to 95\% total variance
    \item Project $\mathbf{x}_0$, $(\mathbf{a}^k_0)_{k\in[1,K]}$ and $(\mathbf{a}^k_t)_{k\in[1,K]}$ on the $n^\mathrm{eof}$ first EOFs
    \item Perform LL analog forecasting in this projected space
\end{itemize}

The second strategy is to perform $n$ analog forecasts, one for each coordinate of the phase-space $\mathcal{P}$, and to assume that the future of a given coordinate only depends on the initial values of the neighboring coordinates and not on the whole initial vector $\mathbf{x}_0$. In the model of \cite{Lorenz1996} (hereafter noted L96), Eq. (\ref{eq:L96}) in appendix motivates the choice of keeping only the initial coordinates $\lbrace i-2, i-1, i, i+1, i+2 \rbrace$ to estimate the $i$-th future coordinate. Thus we keep only $n^\mathrm{trunc}=5$ initial coordinates. Thus, the LL operator performs $n$ linear regressions with 5 coefficients at each forecast. By combining the results of those linear regressions, one finds a $n\times n$ matrix that is sparse by construction: all elements two cells away from the diagonal are equal to zero. This was introduced in \cite{Lguensat2017} as "local analogs". In the present paper this strategy will rather be termed  as "coordinate-by-coordinate" analog forecasting.\medskip

\textbf{Coordinate-by-coordinate forecast}
\begin{itemize}
\small
    \item for $i$ from 1 to $n$,  forecast the $i$-th future coordinate $\mathrm{x}_{t,i}$ :
    \begin{itemize}
        \item Condition the forecast $\Theta_{\mathrm{LL}, i}^t$ on a few initial coordinates around $\mathrm{x}_{0,i}$.\\
        $ \Theta_{\mathrm{LL}, i}^t(\mathbf{x}_0)=\Theta_{\mathrm{LL}, i}^t(\, \mathrm{x}_{0,i-2}\, ,\, \mathrm{x}_{0,i-1}\, ,\, \mathrm{x}_{0,i}\, ,\, \mathrm{x}_{0,i+1}\, ,\, \mathrm{x}_{0,i+2}\, )$
        \item Find analogs of the truncated initial vector  $ (\, \mathrm{x}_{0,i-2}\, ,\, \mathrm{x}_{0,i-1}\, ,\, \mathrm{x}_{0,i}\, ,\, \mathrm{x}_{0,i+1}\, ,\, \mathrm{x}_{0,i+2}\, ) $
        \item Perform LL analog forecasting $\Theta_{\mathrm{LL}, i}^t$
        \item Store the coefficients of the linear regression $ (\, \mathrm{S}_{i,i-2} \, , \mathrm{S}_{i,i-1} \, , \mathrm{S}_{i,i} \, , \mathrm{S}_{i,i+1} \, , \mathrm{S}_{i,i+2} \,)$
    \end{itemize}
    \item Aggregate the coefficients into the $n\times n$ matrix $\mathbf{S}$
\end{itemize}

The next section investigates limitations to the claim that the matrix $\mathbf{S}$ from the LL operator is able to approximate the Jacobian $\boldsymbol{\nabla}\boldsymbol{\Phi}^t$, and studies the impact of dimension reduction techniques on this Jacobian estimation.

\subsubsection{Effect of the number of analogs and the phase-space dimension}

First, to be able to compute $\mathbf{S}$, one must have enough analogs to perform the inversion of the matrix $\mathbf{X}^\mathrm{T}\boldsymbol{\Omega}\mathbf{X}$, where $\mathbf{X}$ is the matrix of the analogs and $\boldsymbol{\Omega}$ the diagonal matrix of the weights. This cannot be done unless $K$, the number of analogs used for the forecast, is superior or equal to $n$, the phase-space dimension. Using the EOF or coordinate-by-coordinate strategies from the previous section, one can reduce the dimension to $n^\mathrm{eof}$ or $n^\mathrm{trunc}$, needing only to satisfy $K\ge n^\mathrm{EOF}$ or $K\ge n^\mathrm{trunc}$.

To illustrate the practical consequences of these issues, numerical simulations of the L96 system were performed with $n=8$. The L96 is a famous chaotic dynamical system with a flexible dimension, well suited to the purpose of this study. The governing equations were solved using a fourth-order Runge-Kutta numerical scheme with an integration time step $\Delta t=0.05$. A catalog was built from one long trajectory (10$^4$ times) using the real equations ($\delta=0$). Then, analog forecasting was performed at lead time 0.05, using the LL operator on $2\times 10^4$ test points ($10^3$ non-dimensional times) taken from another trajectory on the attractor (independent from the catalog). Setting the number of analogs to the limiting case $K=9$ implies that there are just enough analogs to perform the linear regression (plus one extra analog). Even though $n=8$ is not a very large dimension, if one is provided only with 9 good analogs, one must consider dimension reduction. Regular LL analog forecasting was compared with the combination of analog forecasting with EOFs, keeping the EOFs up to 95\% variance, and with the coordinate-by-coordinate analog forecasting, with $n^\mathrm{trunc}=5$.

The EOF strategy ensures that the linear regression can be performed, as it projects the phase-space $\mathcal{P}$ onto the EOFs that maximize the variance in the set of analogs. Thus the rank of the set of analogs is likely to be equal to $n^\mathrm{eof}$ in this reduced-space. However, the EOF strategy necessarily misses some of the components of the full $(n\times n)$ Jacobian matrix $\boldsymbol{\nabla \Phi}^t$, as it gives only the estimation of a $(n^\mathrm{eof}\times n^\mathrm{eof})$ matrix. The coordinate-by-coordinate method also ensures that the linear regression can be performed as long as $n^\mathrm{trunc}$ is low enough, but is also misses some of the elements of the Jacobian matrix of the flow map. Indeed, even though the coefficients of $\boldsymbol{\nabla}\mathbf{f}$ are zero two cells away from the diagonal, this is not the case of $\boldsymbol{\nabla}\boldsymbol{\Phi}^t$. Recall that, at second-order in time, $\boldsymbol{\nabla}\boldsymbol{\Phi}^t = \mathbf{I} + t\, \boldsymbol{\nabla}\mathbf{f} + t^2 \, \left( \left(\boldsymbol{\nabla}\mathbf{f}\right)^2 + \frac{\mathrm{d}}{\mathrm{d}t}\boldsymbol{\nabla}\mathbf{f} \right)$. Thus, some coefficients of order $t^2$ will not be captured by the linear regression matrix $\mathbf{S}$ using coordinate-by-coordinate analog forecasting with $n^\mathrm{trunc}=5$.

\begin{figure}
    \noindent\includegraphics[scale=.63]{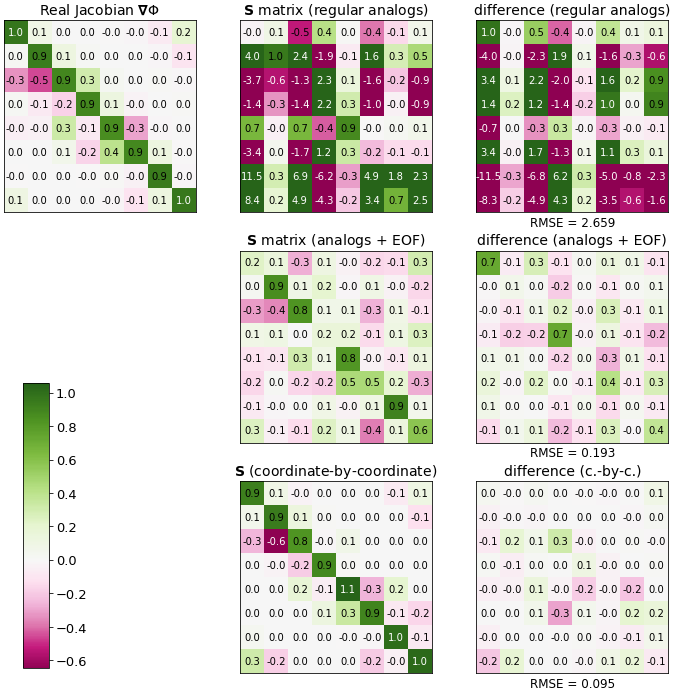}
    \caption{
    Flow map Jacobian matrix estimation with the model of \cite{Lorenz1996}. Forecast lead time is $t=0.05$ Lorenz time, catalog length is 10$^4$ Lorenz times, phase-space dimension is $n=8$. $K=9$ analogs are used for the forecast and Gaussian kernels for the weights $\omega_k$ with shape parameter $\lambda$ set to the median of analog-to-state distances $|\mathbf{a}^k_0-\mathbf{x}_0|$. Upper-left: Jacobian matrix $\boldsymbol{\nabla}\boldsymbol{\Phi}^t|_{\mathbf{x}_0}$. Upper-middle: linear regression matrix $\mathbf{S}$ using regular analogs. Upper-right: difference $\mathbf{S}-\boldsymbol{\nabla}\boldsymbol{\Phi}^t|_{\mathbf{x}_0}$ with regular analogs, also giving the value of RMSE below the plot. Middle panels: same but the linear regression is performed in a lower-dimensional subspace spanned by the first EOFs of the set of the $K=9$ analogs. Lower panels: same but the linear regression is performed coordinate-by-coordinate, and assuming that the coefficients are zero two cells away from the diagonal.
    }
    \label{fig:example_J_L96}
\end{figure}

The linear regression matrix $\mathbf{S}$ is then compared with $\boldsymbol{\nabla}\boldsymbol{\Phi}^t$ for the three methods. The real value of $\boldsymbol{\nabla}\boldsymbol{\Phi}^t$ is estimated with the second-order time-expansion of Eq. (\ref{eq:Phitof_tsquared}) that can be computed directly from the model equations (\ref{eq:L96}). An example is shown in Fig. \ref{fig:example_J_L96}. In this case, the regular analog forecasting misses the Jacobian with RMSE of 2.659, because the rank of the set of analogs is too low and $\mathbf{X}^\mathrm{T}\boldsymbol{\Omega}\mathbf{X}$ is thus not invertible. Analog forecasting combined with EOFs gives a better result as it circumvents this inversion problem, with a total RMSE between $\mathbf{S}$ and $\boldsymbol{\nabla}\boldsymbol{\Phi}^t$ of 0.193. The coordinate-by-coordinate analog forecasting gives the best solution in this case, with a RMSE of 0.095. Note that many coefficients of the matrix $\mathbf{S}$ are set to zero by construction when using the coordinate-by-coordinate method.

\begin{figure}
    \includegraphics[scale=0.6]{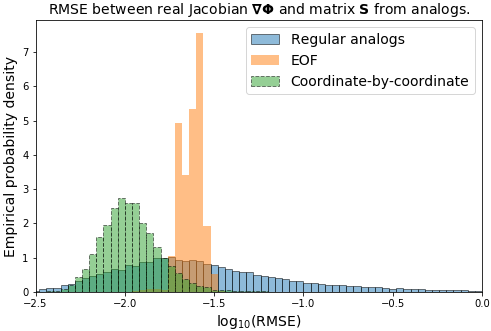}
    \caption{Empirical probability density function of RMSE in flow map Jacobian matrix estimation, depending on the method used. We use the system of \cite{Lorenz1996} with phase-space dimension $n=8$. $K=9$ analogs are used for each forecast and the methods are the same as in Fig. \ref{fig:example_J_L96}. 
    }\label{fig:J_L96_proba}
\end{figure}

Then, Fig. \ref{fig:J_L96_proba} shows empirical probability density functions for the RMSE of $\mathbf{S} - \boldsymbol{\nabla}\boldsymbol{\Phi}^t$ for each of the three methods. The low number of analogs implies large fluctuation of the regular LL analog forecasts, as the rank of the set of analogs used can be below or close to the phase-space dimension, making the inversion of $\mathbf{X}^\mathrm{T}\boldsymbol{\Omega}\mathbf{X}$ hazardous. This variability is noticeably reduced when the inversion is performed in the $n^\mathrm{eof}$-dimension reduced-space. The EOF strategy has the advantage of preventing large errors and the drawback of hindering very precise estimations of the Jacobian. Indeed, when using EOFs the linear regression matrix has a rank necessarily lower than $n$, and some information is missed. Finally, coordinate-by-coordinate analog forecasting is able to perform better estimations of the Jacobian in average, and with a variability between that of the regular analogs and that of the analogs combined with EOFs. However, the probability to have very precise estimations of the Jacobian ($\log_{10}(\mathrm{RMSE})< -2.3$) is lower with coordinate-by-coordinate analog forecasting than with regular analog forecasting. This can be witnessed as the area under the graph for $\log_{10}(\mathrm{RMSE})<-2.4$ is larger for regular analogs then for coordinate-by-coordinate analog forecasting. This is due to the small (order $t^2$) non-zero coefficients two cells away from the diagonal that the coordinate-by-coordinate analog forecasting cannot estimate.\medskip

In some situations however, the number of analogs $K$ is much larger than the phase-space dimension $n$, and the linear regression matrix $\mathbf{S}$ is still unable to approximate the Jacobian $\boldsymbol{\nabla}\boldsymbol{\Phi}^t$.

\subsubsection{Effect of the analogs rank and the attractor's dimension}

As we have seen, to calculate $\mathbf{S}$ and perform locally-linear analog forecasting, one must invert the matrix $\mathbf{X}^\mathrm{T}\boldsymbol{\Omega}\mathbf{X}$. This means that the set of analogs must be of rank $n$. Yet, in some situations, the dimension of the attractor is lower than the full phase-space dimension $n$. Thus if the catalog is made of one trajectory inside the attractor, 
the set of analogs might not be of rank $n$, however large $K$ might be. In some cases, the dimension of the attractor is between $n-1$ and $n$, such that the matrix $\mathbf{X}^\mathrm{T}\boldsymbol{\Omega}\mathbf{X}$ is still invertible but very sensitive to fluctuations in the rank of the set analogs.

Similar remarks can be made for the successors. If $\mathbf{Y}$ (the set of successors) is not of rank $n$, then the matrix $\mathbf{S}$, if it can be computed, is still not of rank $n$. Thus $\mathbf{S}$ will not be able to estimate the Jacobian $\boldsymbol{\nabla}\boldsymbol{\Phi}^t$ if the latter is of rank $n$. Note that the rank of the successors (the rank of the matrix $\mathbf{Y}$) is highly dependent on the rank of the analogs and the Jacobian matrix as we have $\mathbf{Y}\approx\mathbf{X}\boldsymbol{\nabla}\boldsymbol{\Phi}^\mathrm{T}$ at first order in $\mathbf{X}$, such that if the analogs are not of rank $n$ the successors are likely not to be of rank $n$ either.

Thus, depending on the dimension of the attractor, the locally-linear analog forecasting operator might not be able to estimate the local Jacobian of the real flow map, but only a projection of this Jacobian matrix onto the local sets of analogs and successors. This is a typical case where data-driven methods are not able to reveal the full physics of an observed system unless provided with other sources of information or hypotheses, such as a parametric law.

The three-variable L63 system is used to illustrate this fact. This system is known to have a dimension of $\approx 2.06$, with local variations around this value \citep{caby2019generalized}. This is the perfect case study where the rank of the set of analogs will be close to $n-1$. Thus, the linear regression matrix $\mathbf{S}$ between the analogs and the successors is not able to approximate the full $(3\times 3)$ Jacobian matrix $\boldsymbol{\nabla \Phi}^t$. Using restriction to the vector subspace $V_a$ spanned by the two first EOFs of the analogs $(\mathbf{e}^a_1,\mathbf{e}^a_2)$, one can understand better the connection between the two matrices $\boldsymbol{\nabla \Phi}^t$ and $\mathbf{S}$. In the following, subscript "$r$" indicates restriction to $(\mathbf{e}^a_1,\mathbf{e}^a_2)$. The choice of using only the two first EOFs is motivated by the quasi-planar nature of the Lorenz attractor. In the next formulas the $t$-superscript is dropped for the sake of readability.

\begin{subequations}

\begin{align}
    \boldsymbol{\nabla\Phi}_{r} = \boldsymbol{\nabla\Phi} 
    \begin{pmatrix}
        \mathbf{e}^a_1\\
        \mathbf{e}^a_2\\
        0
    \end{pmatrix} \; ,
\end{align}

\begin{align}
    \mathbf{S}_{r} = \mathbf{S} 
    \begin{pmatrix}
        \mathbf{e}^a_1\\
        \mathbf{e}^a_2\\
        0
    \end{pmatrix} \; .
\end{align}

\end{subequations}

The condition number of the set of analogs gives a direct way to measure whether the matrix $\mathbf{X}^\mathrm{T}\boldsymbol{\Omega}\mathbf{X}$ can be inverted, and whether $\mathbf{S}$ can approximate a full rank Jacobian matrix. This number is the ratio of highest to lowest singular value. It has the advantage of being a continuous function of the set of analogs, while the rank is a discontinuous function that takes only integer values.  If the condition number is large, the set of analogs is almost contained in a plane, and the analogs might not be able to approximate the full Jacobian $\boldsymbol{\Phi}$. Conversely, if the condition number is close to 1, then the rank of the set of analogs is clearly 3, and analogs will be able to approximate the full Jacobian matrix. Note that the condition number of the set of analogs is not directly linked to the dimension of the attractor. One simply uses the fact that the attractor is locally close to a plane, without referring further to the complex notion of attractor dimension. 

This can be investigated through numerical simulations of the L63 system, using a fourth-order Runge-Kutta numerical scheme and a time step of $\Delta t=0.01$ to solve the governing equations. A catalog was generated from a trajectory of 10$^5$ non-dimensional times, with the original equations ($\delta=0$). Locally-linear analog forecasting was performed at horizon $t=0.01$ with $K=40$ analogs, on 10$^4$ points randomly selected on the attractor. The linear regression matrix $\mathbf{S}$ was then compared with $\boldsymbol{\nabla}\boldsymbol{\Phi}^t$, with or without restriction to $(\mathbf{e}^a_1,\mathbf{e}^a_2)$. To estimate numerically the real value of $\boldsymbol{\nabla}\boldsymbol{\Phi}^t$, a third-order time-expansion similar to Eq. (\ref{eq:Phitof_tsquared}) was computed directly from the model equations.

\begin{figure}
    \noindent\includegraphics[scale=0.6]{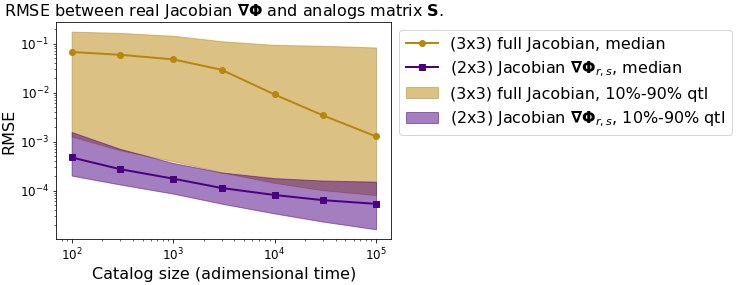}
    \caption{
    RMSE in estimating with analogs the L63 Jacobian matrix, as a function of catalog size. In brown circles, the median RMSE (with 10\% and 90\% quantiles) of the total ($3\times3$) Jacobian matrix. In violet squares, the median RMSE (with 10\% and 90\% quantiles) of the ($2\times2$) Jacobian matrix
    after projection on the two first EOFs of the successors and restriction to the two first EOFs of the analogs. The projection-restriction implies much lower RMSE, and a much lower variability. Both estimation errors are decreasing functions of the catalog size. The number of test points decreases with catalog size, as more test points are needed for small catalogs.}
    \label{fig:J_L63_catalogsize}
\end{figure}

Fig. \ref{fig:J_L63_catalogsize} shows that estimation of the Jacobian by the analogs improves as the catalog size (and therefore the catalog density) grows. This validates that the analogs are able to approximate precisely the Jacobian matrix of the flow map. The figure also shows that, once restricted to the two-dimensional subspace spanned by the analogs, this estimation is much more precise and less fluctuating. \medskip

\begin{figure}
    \noindent\includegraphics[scale=0.6]{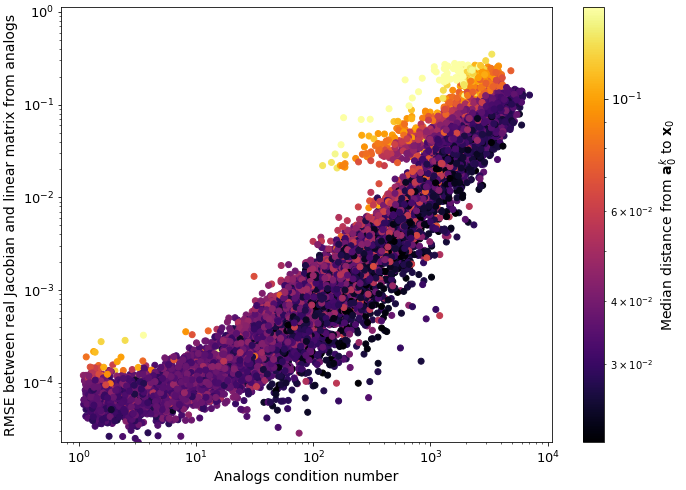}
    \caption{
    RMSE in analogs estimation of the full (3x3) Jacobian matrix $\boldsymbol{\nabla}\boldsymbol{\Phi}^t$ as a function of analog rank and median analog distance, with the L63 system. The rank of each set of analogs is measured by the ratio between the lowest and the highest singular value of the set of analogs. Most of the variability of the RMSE is explained by the rank of the analogs. Some of the remaining variability can be explained by the median distance from the analogs $\mathbf{a}_0^k$ to $\mathbf{x}_0$, which gives a measure of the local catalog density. The catalog size is 10$^5$ non-dimensional times, $\delta=0$, and we use $K=40$ analogs. Tests are done at 10$^4$ points randomly selected on the attractor.
    }
    \label{fig:J_svratio}
\end{figure}

Fig. \ref{fig:J_svratio} displays the RMSE of the full (3$\times$3) matrix $\mathbf{S} - \boldsymbol{\nabla}\boldsymbol{\Phi}$ as a function of the condition number of the set of analogs. We can see in this figure that large RMSE values are highly correlated with high condition numbers, while low RMSE values can only be achieved when the condition number of the analogs is close to 1.

All these elements show that the estimation of the Jacobian matrix from analogs is highly dependent on the number of analogs $K$, the condition number of the set analogs, the attractor's dimension, and the phase-space dimension $n$. However, the fact that the matrix from the LL operator does not approximate the full Jacobian $\boldsymbol{\nabla\Phi}^t$ does not mean that the analog forecast will poorly approximate the future state $\mathbf{x}_t$. For the LL forecast to be efficient, one only needs a good approximation of the restricted Jacobian, and that the inversion associated with the linear regression is not ill-conditioned. 

\subsection{Evolution of mean and covariance under Gaussian assumption} \label{subsec:CovAna}

In this section, it is assumed that the weighted multinomial distribution of the analogs $\sum_k \omega_k\boldsymbol{\delta}_{\mathbf{a}^k_0}$ and of their successors $\sum_k \omega_k\boldsymbol{\delta}_{\mathbf{a}^k_t}$ can be approximated by Gaussian distributions:

\begin{subequations}
\begin{equation}
\sum_k \omega_k\boldsymbol{\delta}_{\mathbf{a}^k_0} \approx \mathcal{N}\left( \boldsymbol{\mu}_0, \mathbf{P}_0 \right) \, ,
\end{equation}
\begin{equation}
\sum_k \omega_k\boldsymbol{\delta}_{\mathbf{a}^k_t} \approx \mathcal{N}\left( \boldsymbol{\mu}_t, \mathbf{P}_t \right) \, ,
\end{equation}
\end{subequations}

\noindent where we have $\boldsymbol{\mu}_t=\boldsymbol{\mu}_\mathrm{LC}$. Combining this hypotheses with Eq. (\ref{eq:DynPhi}b) and approximating $\boldsymbol{\Phi}^t_a(\cdot)$ by its tangent around $\boldsymbol{\mu}_0$ we have the classic relationships:

\begin{subequations}\label{eq:Gauss_muP_F}
\begin{equation}
\boldsymbol{\mu}_t=\boldsymbol{\Phi}_a^t(\boldsymbol{\mu}_0) + \mathcal{O}\left( \mathrm{Tr}\mathbf{P}_0 \right)  \, ,
\end{equation}
\begin{equation}
\mathbf{P}_t=\boldsymbol{\nabla}\boldsymbol{\Phi}_a^t|_{\boldsymbol{\mu}_0} \, \mathbf{P}_0 \, \boldsymbol{\nabla}\boldsymbol{\Phi}_a^t|_{\boldsymbol{\mu}_0}^\mathrm{T} \, + \, \mathcal{O}\left( \mathrm{Tr}\mathbf{P}_0 \right) \, ,
\end{equation}
\end{subequations}

\noindent where $\mathrm{Tr}$ is the trace operator. Similar relations can be found using the differential representation of Eq. (\ref{eq:DynPhi}b):

\begin{subequations}\label{eq:Gauss_muP_f}
\begin{equation}
\dfrac{\mathrm{d}\boldsymbol{\mu}_t}{\mathrm{d}t}=\mathbf{f}_a(\boldsymbol{\mu}_t) + \mathcal{O}\left( \mathrm{Tr}\mathbf{P}_t \right) \, , \quad \boldsymbol{\mu}_{t=0}=\boldsymbol{\mu}_0 \, ,
\end{equation}
\begin{equation}
\dfrac{\mathrm{d}\mathbf{P}_t}{\mathrm{d}t}=\boldsymbol{\nabla}\mathbf{f}_a|_{\boldsymbol{\mu}_t} \, \mathbf{P}_t + \mathbf{P}_t \, \boldsymbol{\nabla}\mathbf{f}_a|_{\boldsymbol{\mu}_t}^\mathrm{T} + \mathcal{O}\left( \mathrm{Tr}\mathbf{P}_t \right)\, , \quad \mathbf{P}_{t=0}=\mathbf{P}_0 \, .
\end{equation}
\end{subequations}

Now, let us make the simplifying hypothesis that $|\mathbf{x}_0-\boldsymbol{\mu}_0|^2\lesssim \mathrm{Tr}\mathbf{P}_0$, which means that the state $\mathbf{x}_0$ is not farther from the analogs' mean $\boldsymbol{\mu}_0$ than the standard deviation of the analogs. Then, one evaluates $\boldsymbol{\Phi}_a^t$, $\mathbf{f}_a$ and their derivatives at $\mathbf{x}_0$ and $\mathbf{x}_t$ instead of $\boldsymbol{\mu}_0$ and $\boldsymbol{\mu}_t$, giving additional terms: 

\begin{subequations}
\begin{equation}\label{eq:Gauss_mu_F}
\boldsymbol{\mu}_t=\boldsymbol{\Phi}^t(\mathbf{x}_0) + \delta \tilde{\boldsymbol{\Phi}}^t(\mathbf{x}_0) + \boldsymbol{\nabla}\boldsymbol{\Phi}^t|_{\mathbf{x}_0}(\boldsymbol{\mu}_0-\mathbf{x}_0) + \mathcal{O}\left( \mathrm{Tr}\mathbf{P}_0 \, , \,  \delta|\boldsymbol{\mu}_0-\mathbf{x}_0| \right)  \, ,
\end{equation}
\begin{multline}\label{eq:Gauss_P_F}
\mathbf{P}_t=\boldsymbol{\nabla}\boldsymbol{\Phi}^t|_{\mathbf{x}_0} \, \mathbf{P}_0 \, \boldsymbol{\nabla}\boldsymbol{\Phi}^t|_{\mathbf{x}_0}^\mathrm{T} \, 
+ \, \delta\left( \boldsymbol{\nabla}\boldsymbol{\Phi}^t|_{\mathbf{x}_0} \, \mathbf{P}_0 \, \boldsymbol{\nabla}\boldsymbol{\Phi}^t|_{\mathbf{x}_0}^\mathrm{T} \,
+ \, \boldsymbol{\nabla}\boldsymbol{\Phi}^t|_{\mathbf{x}_0} \, \mathbf{P}_0 \, \boldsymbol{\nabla}\boldsymbol{\Phi}^t|_{\mathbf{x}_0}^\mathrm{T} \right) \\
+ \left( \left[(\boldsymbol{\mu}_0-\mathbf{x}_0)\boldsymbol{\nabla}^2\boldsymbol{\Phi}^t|_{\mathbf{x}_0}\right]  \mathbf{P}_0  \boldsymbol{\nabla}\boldsymbol{\Phi}^t|_{\mathbf{x}_0}^\mathrm{T} 
+  \boldsymbol{\nabla}\boldsymbol{\Phi}^t|_{\mathbf{x}_0} \mathbf{P}_0 \left[(\boldsymbol{\mu}_0-\mathbf{x}_0)\boldsymbol{\nabla}^2\boldsymbol{\Phi}^t|_{\mathbf{x}_0}\right]^\mathrm{T} \right)  + \mathcal{O}\left( \mathrm{Tr}\mathbf{P}_0 \, , \,  \delta|\boldsymbol{\mu}_0-\mathbf{x}_0| \right) \, ,
\end{multline}
\end{subequations}

\noindent where terms of order $|\boldsymbol{\mu}_0-\mathbf{x}_0|^2$ are included in $\mathcal{O}\left( \mathrm{Tr}\mathbf{P}_0\right)$ and $\boldsymbol{\nabla}^2\boldsymbol{\Phi}^t|_{\mathbf{x}_0}$ is the Hessian of $\boldsymbol{\Phi}^t$ at $\mathbf{x}_0$. In the time-differential representation we have :

\begin{subequations}
\begin{equation}\label{eq:Gauss_mu_f}
\dfrac{\mathrm{d}(\boldsymbol{\mu}_t-\mathbf{x}_t)}{\mathrm{d}t}=\boldsymbol{\nabla}\mathbf{f}|_{\mathbf{x}_t}(\boldsymbol{\mu}_t-\mathbf{x}_t) + \delta \mathbf{\tilde{f}}(\mathbf{x}_t) + \mathcal{O}\left( \mathrm{Tr}\mathbf{P}_t \, , \,  \delta|\boldsymbol{\mu}_t-\mathbf{x}_t| \right)  \, ,
\end{equation}
\begin{multline}\label{eq:Gauss_P_f}
\dfrac{\mathrm{d}\mathbf{P}_t}{\mathrm{d}t}=\boldsymbol{\nabla}\mathbf{f}|_{\mathbf{x}_t} \, \mathbf{P}_t + \mathbf{P}_t \, \boldsymbol{\nabla}\mathbf{f}|_{\mathbf{x}_t}^\mathrm{T} +
+ \, \delta\left( \boldsymbol{\nabla}\mathbf{\tilde{f}}|_{\mathbf{x}_t} \, \mathbf{P}_t \, + \, \mathbf{P}_t  \boldsymbol{\nabla}\mathbf{\tilde{f}}|_{\mathbf{x}_t}^\mathrm{T} \right) \\
+ \left( \left[(\boldsymbol{\mu}_t-\mathbf{x}_t)\boldsymbol{\nabla}^2\mathbf{f}|_{\mathbf{x}_t}\right]  \mathbf{P}_t + \mathbf{P}_t \left[(\boldsymbol{\mu}_t-\mathbf{x}_t)\boldsymbol{\nabla}^2\mathbf{f}|_{\mathbf{x}_t}\right]^\mathrm{T} \right)  + \mathcal{O}\left( \mathrm{Tr}\mathbf{P}_t \, , \,  \delta|\boldsymbol{\mu}_t-\mathbf{x}_t| \right) \, .
\end{multline}
\end{subequations}

Eq. (\ref{eq:Gauss_mu_f}) is equivalent to Eq. (\ref{eq:Gauss_mu_F}), which is also equivalent to Eq. (\ref{eq:Taylor_Ana_Phi}). Eq. (\ref{eq:Gauss_mu_f}) can be Taylor-expanded around $t=0$ to find Eq. (\ref{eq:LCmeanERR}b). This analysis recovers the results from Sec. \ref{sec:ConseqAna}\ref{subsec:MeanAna} for the mean forecast of the locally-constant analog forecasting operator.

Eq. (\ref{eq:Gauss_P_f}) and Eq. (\ref{eq:Gauss_P_F}) are two representations of the same phenomenon. They show that at first order, the growth in covariance between the analogs and successors is directly linked to the Jacobian matrix of $\boldsymbol{\Phi}^t$ at $\mathbf{x}_0$. The covariance of the analog forecast will depend on the covariance of the analogs at $t=0$, $\mathbf{P}_0$, and on the system's local Jacobian $\boldsymbol{\nabla}\boldsymbol{\Phi}^t|_{\mathbf{x}_0}$. 
This is another way to see that the analogs are highly linked to the local dynamics of the system. If the local dynamics induce a large spread in the future possible trajectories, it is captured in the successors' covariance $\mathbf{P}_t$. On the contrary, if the local dynamics are flat ($\boldsymbol{\nabla}\boldsymbol{\Phi}^t|_{\mathbf{x}_0} \simeq \mathbf{I} $ or $\boldsymbol{\nabla}\mathbf{f}|_{\mathbf{x}_0}\simeq \mathbf{0}$) the successors' covariance is equal to the analogs' covariance. \medskip

\section{Discussion}

This paper contributes to the interpretation of analog forecasting methods. Following a similar objective but using different methodology, \cite{Zhao2016} set a mathematical framework for the convergence of analog forecasting operators to the flow map of the real system, with a particular emphasis on the kernels used for the weights $\omega_k$.

There are many natural extensions to the work presented here. The first one is non-deterministic dynamics that can happen, for instance, when forecast is not performed in phase-space but in a lower-dimensional space. One might be provided only with observations of a few variables of the whole system, and try to forecast those same variables. The use of time-embeddings from \cite{Takens1981} combined with analog forecasting is promising \citep{Alexander2017}. Also, \cite{chau2020algorithm} build a catalog of state-space trajectories from a catalog of partial and noisy observations, using analog forecasting and data assimilation.

The second natural extension is to account for observation error in the catalog of analogs. As the flow map is assumed to be quasi-linear in phase-space in the neighborhood of the analogs, one could conduct the same analysis including centered additive noise for each analog and successor of the catalog, and find results similar to the ones outlined here.

One must bear in mind that the use of analog forecasting in applications implies issues such as the choice of the space in which forecasting is performed, the choice of the right metric to compare analogs and initial state, and the combination of analogs with other techniques. In data assimilation, one might want to convert the multinomial distributions of Sec. \ref{subsec:anaop} to Gaussian distributions to use Kalman filtering. Ridge and Lasso regularizations could be used to ease the linear regression instead of the techniques mentioned in Sec. \ref{subsec:commentS}. These operational choices must be made accounting for memory use and computational time (see \citet{Lguensat2017} for differences between regular and coordinate-by-coordinate analog forecasting).

\section*{Conclusion}

Analog forecasting allows to avoid solving complex nonlinear equations by using existing solutions starting from similar initial conditions. The accuracy of analog forecasting depends on local dynamical properties of the system of interest. In particular, the quality of analog forecasts is related to the Jacobian matrix of the real system's flow map, and the linear regression from analogs to successors is shown to provide an approximation of this matrix. This allows to examine the mean accuracy of known analog forecasting operators, and to compare different methods that evaluate this Jacobian matrix, using numerical experiments of famous dynamical systems. The locally-linear operator is found to give the best approximation of the future state, provided that the linear regression is not ill-posed. The locally-incremental operator is shown to give more precise forecasts at small lead times. The Jacobian matrix of the flow map is found to drive the growth of the successors' covariance matrix. Altogether, this brings theoretical evidence that analogs can be used to emulate a real system, and gives quantitative expressions for the precision of analog forecasting techniques.


\acknowledgments
The work was financially supported by ERC grant No. 338965-A2C2 and ANR No. 10-IEED-0006-26 (CARAVELE project).

%
%


%






%
%
%

\appendix[A]

\appendixtitle{Lorenz systems}\label{app:lorenz}

The three-variable "L63"  \cite{Lorenz1963} system of equations is:

\begin{equation}\label{eq:L63}
\begin{cases}
 \dfrac{\mathrm{d}x_1}{\mathrm{d}t}=\sigma(x_2-x_1) \, ,\\ \\
 \dfrac{\mathrm{d}x_2}{\mathrm{d}t}=x_1(\rho-x_3)-x_2 \, ,\\ \\
 \dfrac{\mathrm{d}x_3}{\mathrm{d}t}=x_1x_2-\beta x_3 \, ,
 \end{cases}
\end{equation}

\noindent with usual parameters $\sigma=10$, $\beta=8/3$ and $\rho=28$.

The $n$-variable "L96" \cite{Lorenz1996} system of equations is:

\begin{equation}\label{eq:L96}
\forall i \in [1,n]\, , \quad \dfrac{\mathrm{d}x_i}{\mathrm{d}t} = -(x_{i-2}+x_{i+1})x_{i-1}-x_i+\theta \, ,
\end{equation}

\noindent where $\theta$ is the forcing parameter. We set $n=8$, $\theta=8$, and use periodic boundary conditions $x_{i+n}=x_i$.

\appendix[B]\label{app:hessian}

\appendixtitle{Product of Hessian with vectors}
Let $\mathbf{g}$ a vector-valued, phase-space-dependant function $\mathbf{g}:\mathbb{R}^n\rightarrow\mathbb{R}^n$ such as $\boldsymbol{\Phi}^t$ or $\mathbf{f}$. 

The Hessian of $\mathbf{g}$ at $\mathbf{x}$ is noted $\boldsymbol{\nabla}^2\mathbf{g}|_{\mathbf{x}}$. It is of dimension $n^3$ and its $(i,j,k)$-th coefficient $\left[ \, \boldsymbol{\nabla}^2\mathbf{g}|_{\mathbf{x}} \, \right]_{i,j,k}$ equals $\dfrac{\partial^2 g_k}{\partial x_i\partial x_j}(\mathbf{x})$. The product of a Hessian $\boldsymbol{\nabla}^2\mathbf{g}|_{\mathbf{x}}$ with a $n$-dimensional vector $\mathbf{y}$ is a matrix and its $(i,k)$-th coefficient $\left[ \, \mathbf{y} \boldsymbol{\nabla}^2\mathbf{g}|_{\mathbf{x}} \, \right]_{i,k}$ equals $ \sum_j y_j \dfrac{\partial^2 g_k}{\partial x_i\partial x_j} $. The double-product of a Hessian with two $n$-dimensional vectors $\mathbf{y}$ and $\mathbf{z}$ is a vector and its $k$-th coefficient $\left[ \, \mathbf{y} (\boldsymbol{\nabla}^2\mathbf{g}|_{\mathbf{x}})\mathbf{z}^\mathrm{T} \, \right]_{k}$ equals $ \sum_{i,j} y_j z_i \dfrac{\partial^2 g_k}{\partial x_i\partial x_j} $.  The double product of a Hessian $\boldsymbol{\nabla}^2\mathbf{g}|_{\mathbf{x}}$ with two matrices $\mathbf{X}$ and $\mathbf{Y}$ of same shape $K\times n$ is a matrix of shape $k\times n$ and its $(k,j)$-th coefficient is $\sum_{l,m}X_{k,l}Y_{k,m}\dfrac{\partial^2 g_j}{\partial x_l\partial x_m}$.


\bibliographystyle{ametsoc2014}
\bibliography{biblio}

%

\end{document}